%
\magnification1300  \parskip 2mm plus 1mm \parindent=0pt\def\cl{\centerline}
\def\hs1{\hskip1mm} \def\h10{\hskip10mm} \def\hx{\h10\hbox}
\def\vs{\vskip3mm}
\def\page{\vfill\eject}  
\def\<{\langle} \def\>{\rangle} \def\br{\bf\rm}  \def\it{\tenit}

\def\d{{\rm d}}    
 
\def\A{{\rm A}} \def\B{{\rm B}} \def\X{{\rm X}} \def\Y{{\rm Y}}

\def\ne{=\hskip-3.3mm /\hskip3.3mm}
\def\Pr{{\br Pr}}
\def\Fs{F_{\rm s}} \def\Fsa{F_{{\rm s}1}} \def\Fsb{F_{{\rm s}2}}

\vskip15mm \bf
\cl{Quantum transfer functions, weak nonlocality}
\cl {and relativity} 
{\vskip5mm}
\centerline {by}
\vskip4mm \centerline {Ian C Percival}\vskip6mm 
\cl{Department of Physics} \vskip4mm
\centerline {Queen Mary and Westfield College,
University of London}

\vskip15mm {Abstract}\rm

The method of transfer functions is developed as a tool for studying
Bell inequalities, alternative quantum theories and the associated
physical properties of quantum systems.  Non-negative probabilities
for transfer functions result in Bell-type inequalities.  The method
is used to show that all realistic Lorentz-invariant quantum theories,
which give unique results and have no preferred frame, can be ruled
out on the grounds that they lead to weak backward causality.

\rm

\vfill

98 March 23, QMW TH 98-10/ quant-ph/9803044 \hfill To be submitted to Physics Letters A \eject

\page

\vskip6mm

{\bf 1. Introduction}

In John Bell's  study of the general properties of realistic alternative
quantum theories [B14], he showed that quantum systems have
properties that cannot be simulated by classical systems (References
to Bell's book `Speakable and Unspeakable in Quantum Mechanics' [1]
will be denoted by [B$p$], where $p$ is a page number).  This
discovery led to modern quantum technology and a new branch of quantum
physics.  In order to treat the relativistic version of these
theories,  we need to follow chains of causality.  For this purpose, 
section 2 develops the general theory  of transfer functions that can
be applied to both classical and quantum systems.  The transfer
functions link output events like the results of quantum measurements
to input events like the preparation of quantum systems. 

Section 3 describes the causal relations of classical and quantum
systems in terms of the properties of transfer functions and the
location of the input and output events in spacetime.  Section 4 shows
how the transfer function theory is used to obtain inequalities of the
Bell type, and suggests the use of linear programming.  Section 5 then
uses the transfer function theory to put some very strong constraints
on realistic relativistic quantum theories.

\vs{\bf 2 Transfer functions for classical and quantum systems}

A quantum experiment consists of a quantum system linked to classical
systems through preparation events $i_\A,i_\B,\dots$, which we shall
call inputs, and measurement events $j_\A,j_\B,\dots$, which we call
outputs. The input $i=(i_\A,i_\B,\dots)$ (in the singular) represents
all the input events, and the output $j=(j_\A,j_\B,\dots)$ all the
output events.  Classical systems can similarly be prepared by input
events and measured by output events.

We do not consider as an input or output any event which is 
is the same for all runs of an experiment, for example the production
of a system with total spin zero in a Bell-inequality experiment:
inputs and outputs are variables of an experiment, not constants.
The same event could be both an input and an output.
Input and outputs occupy restricted regions of space and time, and
there is no reason why all the inputs should take place before
all the outputs.  Spacetime relations are discussed in the next
section.

For classical systems without noise, and for some special quantum
experiments, the output $j$ is determined uniquely by the input $i$,
and the relation between them is given by a transfer function $F$ such
that
$$
j = F(i).
\eqno(1)$$ 

If there is a finite number of possible inputs and outputs, then there
is a finite number of possible transfer functions.  For continuous
variations, including continuous preparation and measurement of
quantum systems, there is an infinite number, the mathematics is more
subtle, but the physical principles are the same, so we restrict our
treatment to the finite number.

Suppose for simplicity that the system has only two inputs and
two outputs.  The transfer function then has two components
$$
F = (F_\A,F_\B)\hx{such that}
\h10 j_\A=F_\A(i_\A,i_\B),\h10 j_\B=F_\B(i_\A,i_\B).
\eqno(2)$$

Sometimes the input and output $i_\A,j_\A$
in spacetime region A are far from the input and output $i_\B,j_\B$ in
spacetime region B, and causal relations can the be expressed in
terms of signals between the two regions, as follows:
$$
F_\A(i_\A i_\B)=F_\A(i_\A),\hskip4mm F_\B(i_\A i_\B)=F_\B(i_\B)
\hskip6mm\hbox{(No signals between A and B)}
\eqno(3a)$$$$
F_\A(i_\A i_\B)=F_\A(i_\A),\hskip5mm F_\B(i_\A i_\B)\ne F_\B(i_\B)
\hx{(Signal from A to B only)}
\eqno(3b)$$$$
F_\A(i_\A i_\B)\ne F_\A(i_\A),\hskip5mm F_\B(i_\A i_\B)=F_\B(i_\B)
\hx{(Signal from B to A only)}
\eqno(3c)$$$$
F_\A(i_\A i_\B)\ne F_\A(i_\A),\hskip5mm F_\B(i_\A i_\B)\ne F_\B(i_\B)
\hskip3mm\hx{(Signals both ways),\hskip3mm}
\eqno(3d)$$
where, for example, the inequality in (3b) means that $j_\B$ depends
explicitly on $i_\A$.  Those transfer functions that allow signals are
called {\it signalling} (transfer) functions.  Those that do not are
{\it null}.  The signals are constrained by the conditions of special
relativity, depending on whether the intervals between the inputs and
outputs are spacelike or timelike.

More generally, classical systems are noisy, and the results of
quantum measurements are not determined.  The noise in a classical
system can be represented by a background variable $\lambda$, which
usually represents unknown effects, typically microscopic
fluctuations, which influence the output.  The variable $\lambda$ can
be discrete or continuous, a point in a function space, or any other
space for which a probability distribution $\Pr(\lambda)$ can be
defined [B15].  An example is an electrical circuit with one or more
resistors, where the background variable is a combination of some of
the internal variables of the resistors, whose values are unknown.
The output depends on these background variables, which produce noise
in the resistors.  Bell has given some beautiful examples of
background variables from everyday life [B83,B105,B139,B147,B152].

Suppose that for a given value of $\lambda$, the transfer function
$F_\lambda$ is determined, so that we can express the dependence of
the output on the input in terms of transfer functions $F_\lambda$
such that
$$
j= F_\lambda(i).
\eqno(4)$$
This is just another way of saying the $j$ depends only on the
background variable $\lambda$ and the input $i$, and is uniquely
determined by them.  There are background variables for which the
output is not uniquely determined, but we will suppose that they are
supplemented if necessary by further background variables to make the
output unique for a given $i$ and $\lambda$.

Given the probability $\Pr(\lambda)$ for the background variable, the
probability of the output $j$, given the input $i$, is
$$
\Pr(j/i) = \int\d\lambda\hs1 \Pr(\lambda)\delta(j,F_\lambda(i)),
\eqno(5)$$
where as usual the $\delta$-function is unity when its arguments are
equal and zero otherwise.  For a resistor, $\Pr(\lambda)$ is just the
appropriate Boltzmann distribution.

For a particular system, two values of the background variable
$\lambda$ which have the same transfer function $F_\lambda$ are
completely equivalent.  For a given input they result in the same
output.  Only the function $F_\lambda$ matters.  So all the
information that is needed about the probability distribution
$\Pr(\lambda)$ is contained in the probability distribution
$$ 
\Pr(F) = \int\d\lambda\hs1\Pr(\lambda)\delta(F,F_\lambda).  
\eqno(6)$$ 
For this system, the transfer functions can be treated as if
they were background variables.  The conditional probability for
finding the output $j$, when the input is $i$, is then given by
$$ 
\Pr(j/i) = \sum_F\Pr(F)\delta(j,F(i)).
\eqno(7)$$

The causal relations for a noisy system with background variables can
be expressed in terms of signals just as they were for deterministic
systems in (3).  If the probability $\Pr(F)$ of a signalling transfer
function is not zero, then there are corresponding signals for the
noisy system, but unless the probability is unity, the channel
capacity is less.

For the classical dynamics of an electrical circuit with resistors,
those internal freedoms of the resistors which produce noise in the
circuit are background variables, which are hidden, because we cannot
see the motion of the `classical electrons'.  Unlike some background
variables of quantum mechanics, there is no physical principle that
prevents the hidden variables of the classical circuit from becoming
visible.
 
The same analysis applies to a quantum system prepared in a pure
state, with a very important difference: the background variables are
now hidden variables in principle, and there is no guarantee that they
have the properties of classical background variables.  When a system
does not have these properties, the dependence of the output on the
input must have a quantum component.  A quantum system can do things
that a classical system cannot, which is the basis of technologies
like quantum cryptography and quantum computation.  If the system is
in a black box, an experimenter can tell, just by controlling the
input and looking at the output, that the output is linked to the
input by a quantum system.  This is why Bell's study of hidden
variables has had such important consequences.

Equation (7) for a Bell experiment, together with the condition of
non-negative $\Pr(F)$ for all $F$, requires $\Pr(j/i)$ to satisfy
inequalities, including the Bell inequality.  The method can be used
to find inequalities of the Bell type.  The study of the constraints
on $P(j/i)$ for a general quantum system becomes a problem in linear
programming.

\vs{\bf 3. Spacetime and causality for classical and quantum systems}

Special relativity puts strong constraints on causality in spacetime.

According to classical special relativity, causal influences such as
signals operate at a velocity less than or equal to the velocity of
light, so that an input event influences only output events in or on
its forward light cone, and an output is only influenced by events in
or on its backward light cone.  They cannot operate over spacelike
intervals, nor can they operate backwards in time.  There is only
{\it forward causality}.

Nevertheless, there can be correlations between systems with spacelike
separation, due to background variables that originate in the region
of spacetime common to their backward light cones [2,B55].  Causality
which acts over spacelike intervals is generally called nonlocal
causality, or simply {\it nonlocality}. It does not exist for
relativistic classical systems.  Correlation between events with
spacelike separation is not sufficient evidence for nonlocality, since
it can be due to common background variables.

Causality which acts backwards over timelike intervals means that the
future influences the present, and the present influences the past.
This is {\it backward causality}.  This term will only be used when
there is timelike separation between cause and effect.

These relations can be expressed in terms of inputs and outputs.
Suppose that the input events or inputs $i_\A,i_B,\dots$ and output
events or outputs $j_\A, j_\B,\dots$ are all so confined in spacetime
that the interval between any pair of such events is well-defined as
spacelike or timelike.  The special case of null intervals is
excluded, as it is not needed for our purposes.  Since no signal can
propagate faster than the velocity of light, an output $j_\Y$ depends
only on those $i_\X$ that lie in its backward light cone, so that any
variation of the other $i_\X$ makes no difference to $j_\Y$.  This
condition on the transfer function $F$ is a condition of forward
causality, which applies to all those deterministic classical and
quantum systems for which the input uniquely determines the output.

The background variable $\lambda$ of noisy classical systems may be
considered as an additional input, so it does not affect the causality
relations between the inputs and outputs.  Each of the transfer
functions $F_\lambda$ satisfies the same causality conditions as the
transfer function $F$ of a deterministic system.  There is only
forward causality.

The same applies to the transfer functions $F_\lambda$ of some quantum
systems, but not to all.  According to quantum mechanics, there are
experiments for which the $F_\lambda$ with hidden variables $\lambda$
do not satisfy the conditions of forward causality: causality is
nonlocal.  These include experiments designed to test the violation of
Bell's inequalities.  Experimental evidence has been overwhelmingly in
favour of quantum mechanics, and also tends to favour nonlocal
causality through violation of inequalities, although there are still
some loopholes that need to be closed [3].  Nonlocal causality,
conditional on hidden background variables, is called weak
nonlocality, and the corresponding signals are weak nonlocal signals.
Weak signals require the hidden variables to be known at the receiver,
and since the variables are hidden, the signals are hypothetical.
Nevertheless, the example of Bell experiments shows that the presence
or absence of weak signals can tell us something about the properties
of quantum systems.  Because the background variables are hidden, weak
nonlocal signals cannot be used to send signals faster than the
velocity of light.  However, systems with weak nonlocal signals have
observable properties that cannot be simulated by certain other systems,
such as any system whose inputs and outputs are linked by classical
variables only.

There are is absolutely no experimental evidence for backward
causality of any kind in classical or quantum systems, even weak
backward causality.  We show in section 5, that without weak backward
causality, certain relativistic realistic quantum theories are
impossible.

\vs{\bf 4. Inequalities of the Bell type}

Einstein-Podolsky-Rosen-Bohm experiments, and Bell experiments which
test Bell's or Clauser-Holt-Horne-Shimony (CHHS) [B86] inequalities, are
typical quantum nonlocality thought experiments.  In one run of an
experiment, two spin-half particles with total spin zero are ejected
in opposite directions from a central source.  This stage of the
preparation is not included in the input, as it happens for every run
of the experiment, and the inputs in $i$ are the variable inputs, not
the constant inputs.  Before the particles are detected, they pass
into Stern-Gerlach magnetic fields.  The classical settings of the
orientations of the magnetic fields are the input events.  The
measurements, which include the classical recording of one of the two
spin directions parallel or antiparallel to the field, are the output
events.  Most real experiments have been based on photon polarization,
but the principles are the same.

There are two inputs and two outputs, given by
$$
i=(i_\A i_\B),\h10 j=(j_\A j_\B),
\eqno(8)$$
where one pair of inputs and outputs is confined to a spacetime region
A and the other is confined to a spacetime region B.  The output
$j_\A$ is in the forward light cone of $i_\A$ and similarly for $j_\B$
and $i_\B$.  The input preparation at A must be separated by a
spacelike interval from the output measurement at B and vice versa.
This is a considerable experimental challenge.

For local hidden variables, it is not possible to send even weak
signals between A and B, so there are only transfer functions of type
(3a), and the other 3 types of transfer function have probability zero.
So the only transfer functions that have nonzero probability have the
form
$$
F = (F_\A,F_\B), \hx{where} \hskip5mm
j_\A = F_\A(i_\A),\hskip5mm j_\B = F_\B(i_\B).
\eqno(9)$$
The conditional probability for the outcome $j_\A,j_\B$ given
inputs $i_\A,i_\B$ is then
$$
\Pr(j_\A,j_\B/i_\A,i_\B) = \sum_{F_\A,F_\B}
\Pr(F_\A,F_\B)\delta(j_\A,F_\A(i_\A))\delta(j_\B,F_\B(i_\B)).
\eqno(10)$$
Using $(F_\A,F_\B)$ as a hidden variable $\lambda$ in equations (6)
and (10) results in Bell's [B16], equation (2) for independent
determination of the results at A and B, given $\lambda$.

The spin of a particle is denoted $+$ when it is in the same direction
as the field and $-$ when it is in the opposite direction.  The output
$j=(j_\A j_\B)$ then consists of one of four pairs of spin components,
$(++)$, $(+-)$, $(-+)$ and $(--)$, where the first sign represents the
output at A and the second sign the output at B.  We will adopt a
convention whereby the angles at A and B are measured from zeros that
point in opposite directions, so that, for total spin zero, when the
angles are equal, the measured signs of the spin components are the
same, not opposite.

Consider first an experiment for which each of the magnetic field
orientations at A and at B has only two possible angles, $\theta_i$
with $i=1,2$, the same pair of angles at A and B, with the above
convention.  Then the function $F_\A$ can be labelled by a table of
its output values $(\pm)$ for $i=1$ and $i=2$, and similarly for
$F_\B$.  Thus if $j_\A=+$ for $i_\A = 1$ and $j_\A=-$ for $i_\A=2$, we
can denote $F_\A$ by $F_A=[+-]$.  The function $F$ can then be labelled
by a table of 4 values, those for A first, and those for B second, eg
$[+-,+-]$.  For general functions $F$ we would have $2^4=16$ output
values.  But since we are assuming local hidden variables, $F$ has the
form (9), and only 2 arguments each are needed to specify $F_\A$ and
$F_\B$, so we need only 4 output values to specify $F$.

The number of independent probabilities $\Pr(F)=\Pr(F_\A,F_\B)$ is
then severely restricted by the fact that for opposite input
orientations of the magnetic fields at A and B, corresponding to
$\theta_\A=\theta_\B$ and $i_\A=i_\B$, the output signs $j_\A,j_\B$
must be the same.  So when they are different the probability of $F$
is zero.  Reversing the direction of all spins does not affect the
probabilities.  Thus all $\Pr(F)$ are zero except
$$
P_1 = \Pr([++,++])  = \Pr([--,--]),
\eqno(11a)$$$$
P_2 = \Pr([+-,+-])  = \Pr([-+,-+]),
\eqno(11b)$$
where the right-hand equalities follow by symmetry.  Notice that the
two strings of signs, for A and for B, are the same.

There are no contradictions here: two directions can be handled with
local hidden variables.  But now consider 3 directions
$\theta_1,\theta_2,\theta_3$, as in a Bell inequality.  The values of
$i_\A$ and $i_\B$ are 1,2,3, and there are four independent
probabilities, given by
$$
P_0 = \Pr([+++,+++]) = \Pr([---,---]),
$$$$
P_1 = \Pr([-++,-++]) = \Pr([+--,+--]),
$$$$
P_2 = \Pr([+-+,+-+]) = \Pr([-+-,-+-]),
$$$$ 
P_3 = \Pr([++-,++-]) = \Pr([--+,--+]).
\eqno(12)$$
From these probabilities, we can obtain the known conditional
probabilities for experiments in which $i_\A,i_\B$ with $i=1,2,3$
correspond to the three orientations $\theta_i$ of the fields.
$\Pr(j_\A j_B/i_\A i_\B)$ is given by adding together the
probabilities for which $j_\A$ appears in position $i_\A$ in the first
sequence of signs, and $j_\B$ appears in position $i_\B$ in the second
sequence:
$$
\Pr(++/11) = \Pr(--/11) =  P_0 + P_1 + P_2 + P_3= 1/2 \hx{and cyclic,}
\eqno(13a)$$$$
\Pr(++/23) = \Pr(--/23) = P_0 + P_1  \hx{and cyclic,}
\eqno(13b)$$$$
\Pr(+-/23) = \Pr(-+/23) = P_2 + P_3  \hx{and cyclic,}
\eqno(13c)$$
where the cyclic permutations permute 1,2 and 3 but not 0.

The equations can be solved for the $P_k$, all of which must be
nonnegative.  Their solutions, with resultant conditions on
$\Pr(++/i_\A i_\B)$, $\Pr(+-/i_\A i_\B)$ are
$$
2P_0 = \Pr(++/23)+ \Pr(++/31)+ \Pr(++/12)
-{1\over 2} 
         \ge 0 \hx{and cyclic,}
\eqno(14a)$$$$
2P_1 = {1\over 2}+ \Pr(++/23)- \Pr(++/31)- \Pr(++/12)  
\ge 0  \hx{and cyclic,}
\eqno(14b)$$$$
2P_1 = \Pr(+-/31)+ \Pr(+-/12)- \Pr(+-/23)\ge 0  \hx{and cyclic}.
\eqno(14c)$$
The inequalities (14a,b) are always satisfied, but (14c) gives the
Bell inequalities [B18], his equation (15).  Bell uses the same
origin for the angles at A and B, so his opposite signs are the same
signs here, and his expectation values of products are
$$
P({\bf b},{\bf c}) = 4\Pr(+-/23) - 1  \hx{and cyclic}.
\eqno(15)$$
The condition of weak nonlocality and the conditions that the transfer
function probabilities should all be non-negative, and that
sums over them are all less than or equal to 1,  provide a
systematic method for obtaining inequalities of the Bell type.

Since quantum mechanics violates the Bell inequalities, it is weakly
nonlocal, and there are signalling transfer functions $\Fs$ with
nonzero probabilities $\Pr(\Fs)$.

\vs{\bf 5. Relativistic hidden variables}

It has long been known that it is difficult to reconcile special
relativity and quantum theories that are based on background variables
[B169-172], including the de Broglie-Bohm pilot wave theory [4,5], and
theories based on spontaneous localization or quantum state diffusion
[6-13].  Local `beables', in the sense of Bell [B52] appear to be
incompatible with special relativity.  The difficulty is to reconcile
the weak nonlocality of measurement processes with a theory that
provides a unique Lorentz-invariant result for a measurement when
there is no preferred frame, as discussed in [4].

By taking Einstein's old simultaneity thought experiment using light
signals [14], and replacing his central source and two receivers by a
Bell experiment, the following shows that any relativistic hidden
variable theory with unique results and no preferred frame results in
weak backward causality, and can be rejected for this reason.  If we
exclude backward causality, then relativistic hidden variable quantum
theories of measurement must depend on environmental or cosmological
influences, which define a special frame.

Here is the argument.  It does not require weak locality.  Because of
the Bell inequalities, it must allow weak nonlocal signals.

A Bell inequality experiment on spin one-half particles has a source
S$_1$ and two receivers A$_1$ and B$_1$ at equal distances from S$_1$,
all at rest in the same frame, number 1.  This experiment is
symmetrical with respect to interchange of A$_1$ and B$_1$.  Bell's
theorem shows that, given the hidden variable $\lambda_1$, the result
of the experiment at B$_1$ is dependent on the angle of measurement at
A$_1$, as in (3b), {\it or} the result of the experiment at A$_1$ is
dependent on the angle of the measurement at B$_1$, as in (3c), {\it
or} both.  Now because of symmetry with respect to interchange of A
and B, it {\it must} be both.  There are weak signals in both
directions, with the same probabilities.

Let $\Fs$ with nonzero probability $\Pr(\Fs)$ be one of the signalling
transfer functions.  Then for $j=\Fs(i)$, there must be at least
one value of $i_\B$, defining the orientation at B, and two values
$i_\A=1,2$ which then result in different signs for $j_\B$.  Otherwise
there could be no weak signal.  If we fix the value of $i_\B$ and
allow $i_\A$ to have those two values, $\Fs$ defines a one-bit weak
signal, which is sent with probability $\Pr(\Fs)$ in the Bell experiment.

Now choose two such Bell experiments which are in the same line, but
are at rest in different Lorentz frames with a nonzero relative
velocity parallel to the line of the experiments.  Let the second
frame, with its source and two receivers, be labelled by 2.  We use
$\lambda_2$ to represent the hidden variable that affects the result
of experiment 2.  As in Einstein's experiment, A$_1$ is in the past
light cone of A$_2$ but B$_1$ is in the future light cone of B$_2$.
It is first assumed that the results of the experiments are not
correlated through the hidden variables, so that
$$
\Pr(\lambda_1,\lambda_2) = \Pr(\lambda_1) \Pr(\lambda_2).  
\eqno(16)$$
Later we will drop this assumption.  

There is a weak channel from B$_1$ to A$_1$ and another from
A$_2$ to B$_2$. 

Set up the apparatus with a classical signal from A$_1$ to A$_2$ such
that results of the measurements of A$_1$ determine the orientation of
the measurement at A$_2$.  For example, we could set up the apparatus
at A so that if the result of the measurement at A$_1$ is $+$, then
the orientation of A$_2$ is given by $i_\A =1$, and if the result
of the measurement is $-$, then the orientation of A$_2$ is given by
$i_\A=2$.

Given $\lambda_1$ and $\lambda_2$, the orientation of B$_1$ affects
the result of measurement A$_1$, which determines the orientation of
A$_2$, which affects the result of measurement B$_2$.  Thus the
orientation of B$_1$ affects the sign of the spin at B$_2$.  This
means that weak causality goes backwards in time: there is weak
backward causality.

It would indeed be remarkable if the assumption (16) were false, so
that the correlation of $\lambda_1$ and $\lambda_2$ produced a
correlation between the outputs $j$ for the two Bell experiments.
Then two Bell experiments set up with the spacetime configuration of
Einstein's simultaneity thought experiment, but otherwise unconnected,
would have correlated outputs, which could be checked experimentally.
This is not inconsistent with some alternative quantum theories, like
primary state diffusion [12,13].

However, our conclusion does {\it not} depend on (16).

From section 2, the only way that the correlation of $\lambda_1$ and
$\lambda_2$ could affect the results of the experiments is through
the transfer functions, and since it is the signalling 
functions that produce the backward causality, they would have
to be correlated:
$$
\Pr(\Fsa,\Fsb) \ne \Pr(\Fsa)\Pr(\Fsb).
\eqno(17)$$
This could be achieved if the probability $\Pr(\Fsa,\Fsb)$ of the
two signalling functions occurring together was zero.  In that case a
signal from B1 to A1 and from A2 to B2 would never occur together, so
there would be no backward causality.  But in that case the total
probability of either $\Fsa$ or $\Fsb$ would have to be the sum
of the probabilities, which are equal, so that, since no probability
can be greater than 1:
$$
\Pr(\Fsa{\hs1{\rm or}\hs1}\Fsb)= 2\Pr(\Fsa)\le 1.
\eqno(18)$$

Since $\Pr(\Fsa)$ is small, this inequality can be
satisfied.  But the probability $\Pr(\Fsa)$ cannot be
zero, since there is weak nonlocality.  Suppose that $N$ is the
smallest integer for which
$$
\Pr(\Fsa) > {1\over 2N+1}
\eqno(19)$$
Now replace the two Bell experiments by $2N+1$ of them, in which
each pair has the spacetime overlap property of the original two.
This can be achieved by choosing the preparation events at points
which are distance $L$ apart in some frame:
$$
(t_k,x_k,y_k,z_k) = (\mp L\sinh k\phi,\pm L\cosh k\phi,0,0)
\h10 (k=-N,\dots,N),
\eqno(20)$$
where the frames correspond to boosts of $v=\tanh k\phi$ relative to a
standard Lorentz frame, with $c=1$.  The measurement events take place
after a time interval $\tau$ small compared with $L\sinh\phi$ in the
appropriate frame.

If there is to be no weak signal for {\it any} pair, then none of 
the weak signals can occur together, so
$$
(2N+1)\Pr(\Fsa)\le 1,
\eqno(21)$$ 
which contradicts the inequality (19).  So at least two of the Bell
experiments have a finite probability of their signalling functions
occurring together, resulting in backward causality.  So it is not
possible to avoid backward causality even by allowing the hidden
variables of the Bell experiments to be correlated.

\page{\bf 6. Conclusions}

Transfer functions and their probabilities are a powerful tool for
studying hidden variable theories and the associated physical
properties of quantum systems.  Inequalities of the Bell type follow
from the condition that all transfer functions have positive or zero
probabilities, which suggests the use of the methods of linear
programming.  Transfer function analysis also shows that an important
type of relativistic Lorentz-invariant hidden variable theory can be
ruled out on the grounds that it leads to backward causality.

\vs{\bf Acknowledgments} I thank Lajos Di\'osi for a stimulating
communication, David Dunstan, Nicolas Gisin, William Power, Tim
Spiller, Walter Strunz and particularly Lucien Hardy for very helpful
comments, and also the Leverhulme Foundation and the UK EPSRC for
financial support.
 
\vskip6mm

{\bf References}

[1] J.S. Bell. Speakable and Unspeakable in Quantum Mechanics',  Cambridge,
 University Press, 1987.

[2] O. Penrose and I.C. Percival {\it Proc. Phys. Soc.} {\bf 79} 605, 1962.

[3] E.S. Fry and T. Walter in {\it Experimental Metaphysics} (eds R.S.
Cohen, M. Horne and J. Stachel, Dordrecht) Kluwer Academic, 1997, p91.

[4] D. Bohm and B.J. Hiley {\it The Undivided Universe} London,
Routledge, 1993.

[5] P.R. Holland {\it The Quantum Theory of Motion} Cambridge,
University Press, 1993.

[6] P. Pearle. {\it Phys. Rev. D.} {\bf 13} 857-868, 1976.

[7] N. Gisin. {\it Phys. Rev. Lett.} {\bf 53} 1775-1776, 1984.

[8] G.-C. Ghirardi, A. Rimini and T. Weber. {\it Phys. Rev. D}
{\bf 34} 470-491, 1986.

[9] L. Di\'osi. {\it Phys. Lett. A} {\bf 114} 451, 1986.

[10] N. Gisin. {Helv. Phys. Acta} {\bf 62} 363-371, 1989.

[11] P. Pearle. {Phys. Rev. A} {\bf 39} 2277-2289, 1989.

[12] I.C. Percival {\it Proc. Roy. Soc. A} {\bf 451} 503-513, 1995.

[13] I.C. Percival and W. Strunz {\it Proc. Roy. Soc. A} {\bf 453}
431-446, 1997.

[14] A. Einstein.  {\it Ann. d. Physik} {\bf 17} 891-921, 1905.

\end